\documentclass{aa}

\usepackage{graphicx}
\usepackage[varg]{txfonts}
\usepackage{natbib}
\usepackage[pdftex,colorlinks=true,linkcolor=blue,citecolor=blue,urlcolor=blue]{hyperref}

\graphicspath{{./figures/}}

\begin{document}
\title{The HARPS search for southern extra-solar planets}
\subtitle{XLIII. A compact system of four super-Earth planets orbiting HD~215152
\thanks{The HARPS data used in this article are available in electronic form
at the CDS via anonymous ftp to cdsarc.u-strasbg.fr (130.79.128.5)
or via \url{http://cdsweb.u-strasbg.fr/cgi-bin/qcat?J/A+A/}.}}
\titlerunning{HD~215152 system}
\author{J.-B. Delisle\inst{1,2}
  \and D.~S\'egransan\inst{1}
  \and X.~Dumusque\inst{1}
  \and R.F.~Diaz\inst{1,3,4}
  \and F.~Bouchy\inst{1}
  \and C.~Lovis\inst{1}
  \and F.~Pepe\inst{1}
  \and S.~Udry\inst{1}
  \and R.~Alonso\inst{5,6}
  \and W.~Benz\inst{7}
  \and A.~Coffinet\inst{1}
  \and A.~Collier Cameron\inst{8}
  \and M.~Deleuil\inst{9}
  \and P.~Figueira\inst{10}
  \and M.~Gillon\inst{11}
  \and G.~Lo Curto\inst{12}
  \and M.~Mayor\inst{1}
  \and C.~Mordasini\inst{7}
  \and F.~Motalebi\inst{1}
  \and C.~Moutou\inst{9,13}
  \and D.~Pollacco\inst{14}
  \and E.~Pompei\inst{12}
  \and D.~Queloz\inst{1,15}
  \and N.C.~Santos\inst{10,16}
  \and A.~Wyttenbach\inst{1}
}
\institute{
  Observatoire de l'Université de Genève, 51 chemin des Maillettes, 1290, Sauverny, Switzerland\\
  \email{jean-baptiste.delisle@unige.ch}
  \and 
  ASD, IMCCE, Observatoire de Paris - PSL Research University, UPMC Univ. Paris 6, CNRS,\\
  77 Avenue Denfert-Rochereau, 75014 Paris, France
  \and 
  Universidad de Buenos Aires, Facultad de Ciencias Exactas y Naturales, Buenos Aires, Argentina
  \and 
  CONICET - Universidad de Buenos Aires, Instituto de Astronom\'ia y F\'isica del Espacio (IAFE), Buenos Aires, Argentina
  \and 
  Instituto de Astrof\'isica de Canarias, 38205 La Laguna, Tenerife, Spain,
  \and 
  Dpto. de Astrof\'isica, Universidad de La Laguna, 38206 La Laguna, Tenerife, Spain
  \and 
  Physikalisches Institut, Universitat Bern, Silderstrasse 5, CH-3012 Bern, Switzerland
  \and 
  School of Physics and Astronomy, University of St Andrews, North Haugh, St Andrews, Fife KY16 9SS
  \and 
  Aix Marseille Universit\'e, CNRS, LAM (Laboratoire d'Astrophysique de Marseille) UMR 7326, 13388, Marseille, France
  \and 
  Instituto de Astrof\'isica e Ci\^encias do Espa\c{c}o, Universidade do Porto, CAUP, Rua das Estrelas, 4150-762 Porto, Portugal
  \and 
  Institut d'Astrophysique et de G\'eophysique, Universit\'e de Li\`ege, All\'ee du 6 Ao\^ut 17, Bat. B5C, 4000, Li\`ege, Belgium
  \and 
  European Southern Observatory, Karl-Schwarzschild-Str. 2, D-85748 Garching bei M\"unchen, Germany
  \and 
  Canada France Hawaii Telescope Corporation, Kamuela, 96743, USA
  \and 
  Department of Physics, University of Warwick, Coventry, CV4 7AL, UK
  \and 
  Cavendish Laboratory, J J Thomson Avenue, Cambridge, CB3 0HE, UK
  \and 
  Departamento de F\'isica e Astronomia, Faculdade de Ci\^encias, Universidade do Porto, Rua do Campo Alegre, 4169-007 Porto, Portugal
}

\date{\today}

\abstract{
  We report the discovery of four super-Earth planets around \object{HD~215152},
  with orbital periods of 5.76, 7.28, 10.86, and 25.2~d, and minimum
  masses of 1.8, 1.7, 2.8, and 2.9~$M_\oplus$ respectively.
  This discovery is based on 373 high-quality radial velocity measurements taken by HARPS over 13 years.
  Given the low masses of the planets, the signal-to-noise ratio is not sufficient to constrain the planet eccentricities.
  However, a preliminary dynamical analysis suggests that eccentricities should be typically lower than about $0.03$ for the system to remain stable.
  With two pairs of planets with a period ratio lower than 1.5, with short orbital periods,
  low masses, and low eccentricities,
  \object{HD~215152} is similar to the very compact multi-planet systems found by \textit{Kepler},
  which is very rare in radial-velocity surveys.
  This discovery proves that these systems can be reached with the radial-velocity technique,
  but characterizing them requires a huge amount of observations.
}

\keywords{planetary systems -- planets and satellites: general -- planets and satellites: detection -- planets and satellites: dynamical evolution and stability}

\maketitle


\section{Introduction}
\label{sec:introduction}

The properties of extrasolar planetary systems are very diverse
and often very different from those in the solar system
\citep[e.g.,][]{mayor_doppler_2014}.
Transit surveys, and especially the \textit{Kepler} telescope,
have discovered many systems of super-Earth planets
in compact orbital configurations.
In particular, \textit{Kepler} found many systems with period ratios
($P_\mathrm{out}/P_\mathrm{in}$) lower than 1.5
\citep{lissauer_architecture_2011,fabrycky_architecture_2014}.
These compact planetary systems raise several interesting questions regarding
their formation and long-term evolution, such as whether these systems formed in situ or
by converging migration of the planets in the proto-planetary disk,
and the role played by mean-motion resonances
in setting up their architecture and in their long-term stability.
These questions are still open,
and new observational data would help in better
constraining the formation models.
While compact systems are favored by observational biases in transit surveys
\citep[e.g.,][]{steffen_period_2015},
such systems are particularly challenging to separate
with the radial velocity (RV) method.
To our knowledge,
only three pairs of planets
with $P_\mathrm{out}/P_\mathrm{in} < 1.5$
have been found using the RV technique:
\begin{itemize}
  \item \object{HD~5319}b and c, giant planets
  with $P_\mathrm{out}/P_\mathrm{in} = 1.38$
  \citep[see][]{robinson_jovianmass_2007,giguere_newly_2015},
  \item \object{HD~200964}b and c, giant planets
  with $P_\mathrm{out}/P_\mathrm{in} = 1.34$
  \citep[see][]{johnson_retired_2011},
and \item \object{GJ~180}b and c, super-Earth planets
  with $P_\mathrm{out}/P_\mathrm{in} = 1.40$
  \citep[see][]{tuomi_bayesian_2014}.
\end{itemize}
Only one of these systems consists of super-Earth planets (\object{GJ~180}).
Several other systems have been claimed, but were subsequently contested
\citep[e.g.,][]{diaz_harps_2016}.

In this article we present a system of four super-Earth planets around
\object{HD~215152}, with periods of 5.76, 7.28, 10.86, and 25.2~d
(period ratios of 1.26, 1.49, and 2.32).
Such a compact system of super-Earth planets
is very similar to the compact systems found by \textit{Kepler}.
Separating these signals required
a huge observational effort, in which 373 high-quality HARPS RV measurements
were taken over 13 years.

In Sect.~\ref{sec:stardata} we describe the stellar properties and
the data we used.
We then describe the method we employed to analyze the data
in Sect.~\ref{sec:analysis}.
In Sect.~\ref{sec:dynamics} we assess the orbital stability
of the system and deduce additional constraints on the orbital parameters.
Finally, we discuss our results in Sect.~\ref{sec:discussion}.

\section{HARPS data and stellar properties}
\label{sec:stardata}

\begin{table}
  \begin{center}
    \caption{Stellar properties of \object{HD~215152}.}
    \begin{tabular}{cc|c}
      \hline
      \hline
      Parameter & (units) & \\
      \hline
      $M$ & ($M_\odot$) & $0.77 \pm 0.015$\\
      $R$ & ($R_\odot$) & $0.73 \pm 0.018$\\
      Spectral type & & K3V\tablefootmark{1}\\
      $P_\mathrm{rot}$ & (d) & 42\\
      $\log R'_{HK}$ & & -4.86\\
      $v \sin i$ & ($\mathrm{km/s}$) & 3.35\tablefootmark{2}\\
      $T_\mathrm{eff}$ & (K) & $4935 \pm 76$\tablefootmark{3}\\
      $\mathrm{[Fe/H]}$ & & $-0.10 \pm 0.04$\tablefootmark{3}\\
      $M_V$ & (mag) & 6.45\tablefootmark{3}\\
      $B-V$ & (mag) & 0.99\tablefootmark{4}\\
      $\pi$ & (mas) & $46.24 \pm 0.26$\tablefootmark{5}\\
      $\log g$ & (cgs) & 4.40\tablefootmark{3}\\
      \hline
    \end{tabular}
    \tablefoot{
      \tablefoottext{1}{\citet{gray_contributions_2003}},
      \tablefoottext{2}{\citet{martinez-arnaiz_chromospheric_2010}}.
      \tablefoottext{3}{\citet{sousa_spectroscopic_2008}},
      \tablefoottext{4}{\citet{hog_tycho_2000}},
      \tablefoottext{5}{\citet{gaia_gaia_2016}}
    }
    \label{tab:I}
  \end{center}
\end{table}

\begin{figure}
  \centering
  \includegraphics[width=\linewidth]{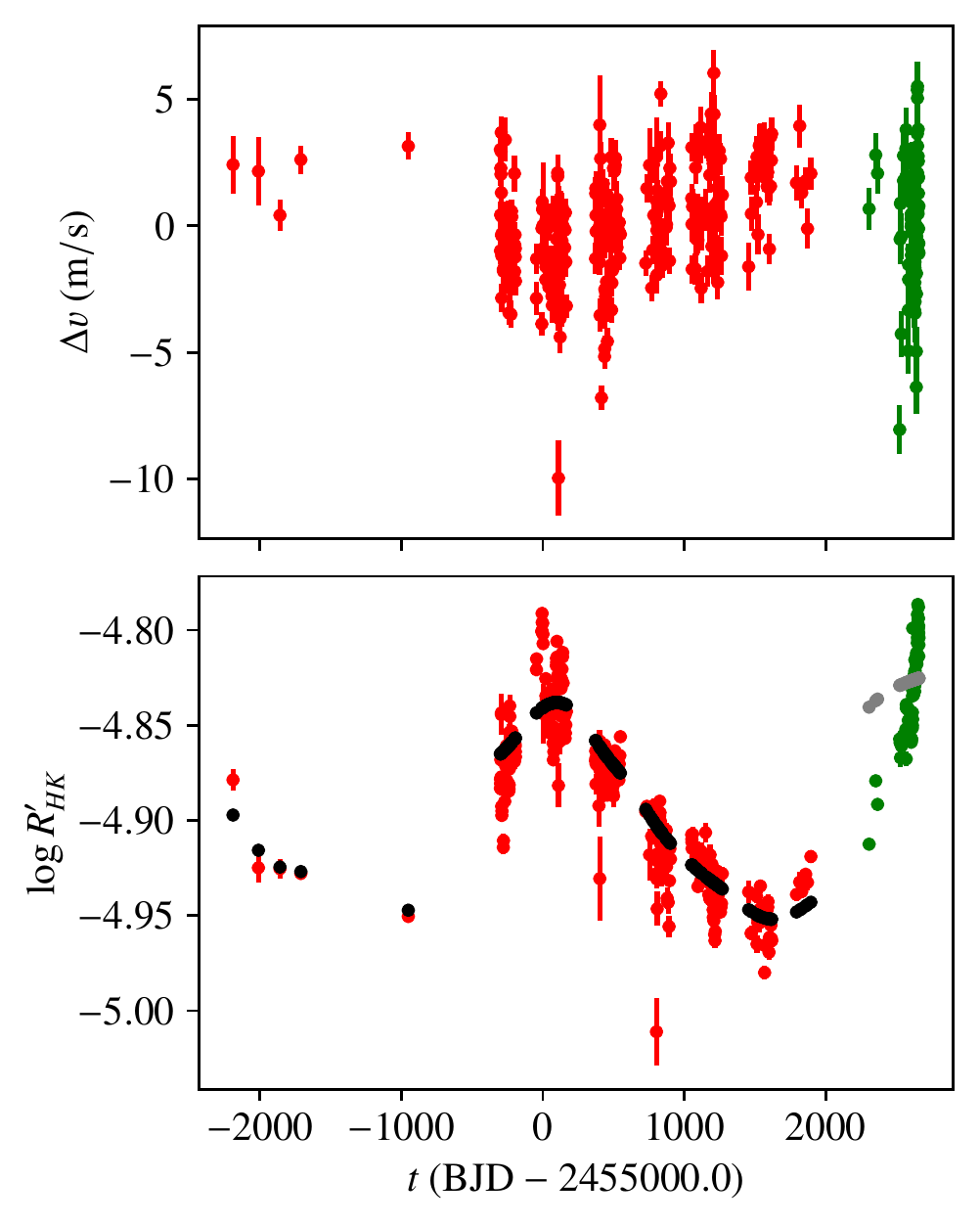}
  \caption{Time series of the RV (\textit{top})
    and the $\log R'_{HK}$ activity index (\textit{bottom}).
    \textit{Red} (\textit{green}) dots correspond to HARPS data
    taken before (after) the fiber upgrade.
    The smoothed activity index,
    described in Sect.~\ref{sec:magcycle},
    is superimposed (\textit{black} and \textit{gray} dots)
    over the raw $\log R'_{HK}$ values for comparison.
    We observe a long-term evolution in the RVs
    that is anticorrelated with the magnetic cycle
    (long-term evolution of the $\log R'_{HK}$).}
  \label{fig:I}
\end{figure}

We present here 373 RV measurements of the K-dwarf \object{HD~215152}
that have been obtained with the HARPS
high-resolution spectrograph mounted on the 3.6m ESO telescope at La Silla Observatory
\citep{mayor_setting_2003}.
The observations span 13 years, from June 2003 to September 2016.
The optical fibers feeding the HARPS spectrograph were changed in May 2015
\citep{locurto_harps_2015}, which results in an offset
between the 284 measurements taken before
and the 89 measurements taken after the upgrade.
The RV points have been taken in 330 different nights.
Of these 330 nights, 287 have a single point,
and 43 have two points.

\citet{dumusque_characterization_2015} found that the stitching of the HARPS detector could introduce
a spurious RV signal of small amplitude with periods close to
one year or an harmonic of it.
The stitching of the detector creates pixels of different sizes
every 512 pixels in the spectral direction,
which perturbs the wavelength solution that is obtained
by interpolating the wavelength of the thorium emission lines on an even scale.
Owing to the barycentric Earth RV,
the spectrum of a star is shifted on the CCD in the spectral direction,
and some lines can cross stitches.
When this is the case, the mismatch in the wavelength solution
at the position of the stitch creates
a spurious RV effect, which, coupled with the shift of
the spectrum on the CCD over time, creates
a signal typically with the period of the rotation of Earth
around the Sun or an harmonic of it.
To remove this spurious signal from the data,
we adopted the method proposed in \citet{dumusque_characterization_2015},
that is, we removed all the lines
that crossed stitches from the computation of the RV .

In addition to the RVs, we also derived
the $\log R'_{HK}$ activity index from each of the 373
spectra.
As for the RV, this index might also be affected by the fiber upgrade.
The RV and $\log R'_{HK}$ measurements are shown in Fig.~\ref{fig:I}.
We also provide a brief summary of the properties of
\object{HD~215152} in Table~\ref{tab:I}.
The mass and the radius of the star as well as their uncertainties
were derived using the Geneva stellar evolution models
\citep{mowlavi_stellar_2012}.
The interpolation in the model grid was made through a Bayesian formalism
using observational Gaussian priors on $T_\mathrm{eff}$,
$M_V$, $\log g$, and [Fe/H] \citep{marmier_phd_2014}.
The rotation period of the star,
estimated using \citet{mamajek_improved_2008}
and \citet{noyes_rotation_1984}, is 42 days.
\object{HD~215152} is a quiet star similar to the Sun,
with a mean $\log R'_{HK}$ of $-4.86$ and
values between $-5$ and $-4.8$
(see Table~\ref{tab:I} and Fig.~\ref{fig:I}).
However, we clearly observe a long-term evolution
in the $\log R'_{HK}$ activity index, with a period of about 3,000~d
(see Fig.~\ref{fig:I}).
This evolution is most likely related to the magnetic cycle of the star.
We also observe a long-period signal in the RVs,
which is anticorrelated with the signal found in the activity index (see Fig.~\ref{fig:I}).
Using only the data taken before the fiber upgrade (to avoid offset issues),
we find a correlation coefficient of
$-0.44 \pm 0.05$
\citep[computed using][]{figueira_pragmatic_2016}
between the RVs and the $\log R'_{HK}$ index.
This anticorrelation
is surprising since a correlation is usually found (and expected)
on the timescale of the magnetic cycle
\citep{lovis_harpsb_2011,dumusque_harps_2011}.
However, for late-K and M stars,
the correlation might reverse
\citep[Fig.~18]{lovis_harpsb_2011},
which might be evidence
for convective redshift \citep{kurster_lowlevel_2003}.

\section{Analysis}
\label{sec:analysis}

We modeled the RVs of \object{HD~215152} as follows:
For the $k$-th data point, we have
\begin{equation}
  \label{eq:model}
  v(t_k) = v_\mathrm{Kep}(t_k) + \gamma_{i}
  + \mathrm{Mag.\ cycle}(t_k) + \epsilon_k,
\end{equation}
where
$v_\mathrm{Kep}(t_k)$ is the Keplerian motion of the star,
$\gamma_{i}$ is the radial proper motion of the system
plus instrumental offset (one independent constant for each instrument),
$\mathrm{Mag.\ cycle}(t_k)$ is the component of the RV generated
by the magnetic cycle of the star, and
$\epsilon_k$ is the noise.
We here only worked with a single instrument (HARPS),
but since the fiber upgrade of May 2015 might induce an offset,
we considered measurements before and after the fiber upgrade as coming from two
independent instruments, and introduced two constants
$\gamma_1$ (before) and $\gamma_2$ (after).
The noise term $\epsilon_k$ in Eq.~(\ref{eq:model})
takes
the white and the correlated components of the noise into account
(instrumental error, short-term activity of the star, etc.).

\subsection{Magnetic cycle correction}
\label{sec:magcycle}

\begin{figure}
  \centering
  \includegraphics[width=\linewidth]{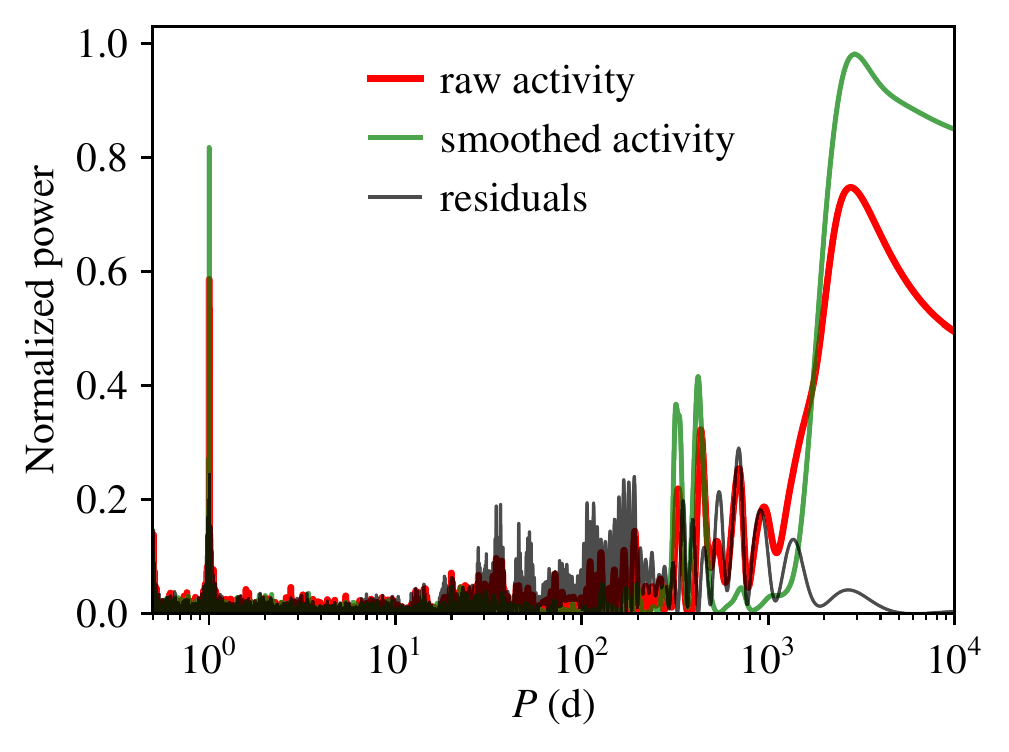}
  \caption{Periodogram of the raw activity ($\log R'_{HK}$ index, in \textit{red}),
    the smoothed activity (see Sect.~\ref{sec:magcycle}, in \textit{green}),
    and the residuals of the activity after the smooth component is subtracted (in \textit{gray}).
    The smoothed activity captures the long-term evolution of the
    $\log R'_{HK}$ index and reduces the high-frequency part of the signal.}
  \label{fig:II}
\end{figure}

\begin{figure}
  \centering
  \includegraphics[width=\linewidth]{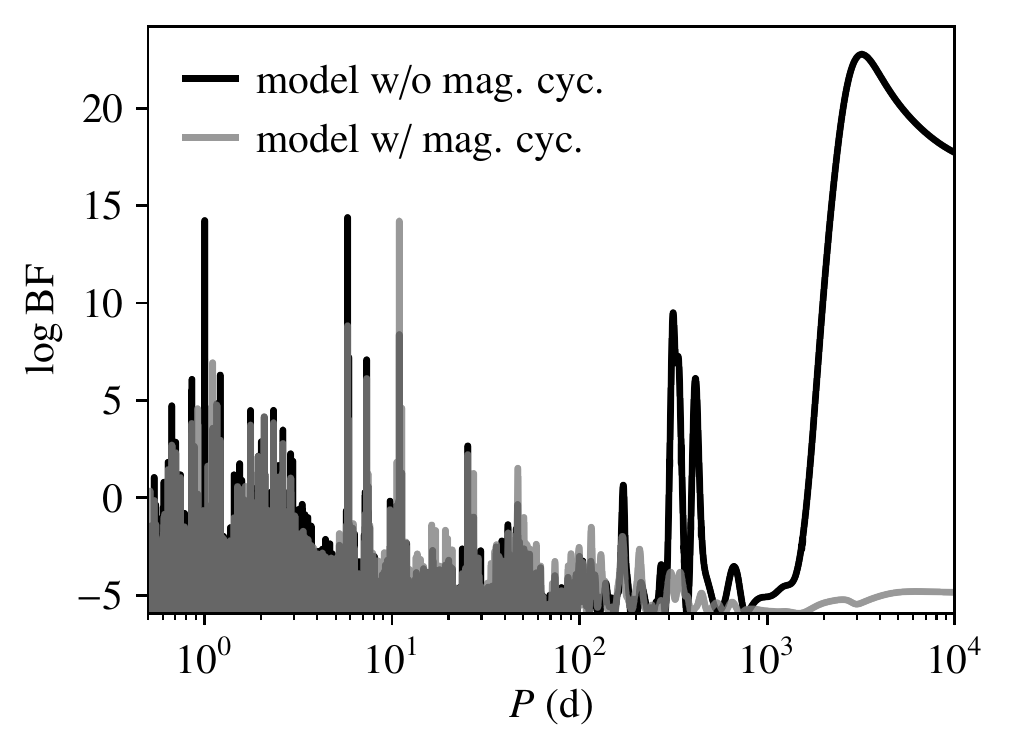}
  \caption{Periodogram of the RVs using two different models.
    The first (\textit{black}) does not include the corrections
    for magnetic cycles,
    while the second (\textit{gray}) does includes them (see Sect.~\ref{sec:magcycle}).
    The main peak in the \textit{black} curve is around 3,000~d
    (similar to the period observed in the $\log R'_{HK}$).
    This peak disappears when the magnetic cycle is corrected for (\textit{gray} curve).
  }
  \label{fig:III}
\end{figure}

In order to account for the effect of the magnetic cycle on the RVs
(see Sect.~\ref{sec:stardata}),
we introduce a proportionality coefficient between the RVs
and the activity index in our model.
While an (anti)correlation is expected between the two quantities
on long timescales because of the magnetic cycles,
on shorter timescales, other effects might break this correlation
(stellar rotation, appearing and disappearing active regions,
or different types of active regions).
In order to avoid introducing artificial short-term signals in the RVs,
we first filtered the $\log R'_{HK}$ index to keep only the long-term trend by
averaging it using a Gaussian kernel.
We used a six-month width for the Gaussian kernel,
which approximately corresponds to one observational season,
in order to clean the high-frequency part of the $\log R'_{HK}$
while preserving the 3,000~d cycle.
Since there might be a small offset in the $\log R'_{HK}$ values
after the fiber change,
we filtered the two data sets independently.
The smoothed activity index is shown in Fig.~\ref{fig:I},
superimposed with the initial $\log R'_{HK}$ index.
In Fig.~\ref{fig:II} we also compare the periodogram
of the raw $\log R'_{HK}$ index,
the smoothed activity index,
and the residuals of the $\log R'_{HK}$ after subtracting this smooth component.
For these periodograms, we adjusted for a possible offset between the data
taken before and after the fiber upgrade.
We observe, as expected, that the smoothed activity index captures the long-term evolution
of the $\log R'_{HK}$ index while cleaning the high-frequency part of the signal (see Fig.~\ref{fig:II}).
We do not observe any significant signal
at the rotation period of the star (around 42~d)
in the periodogram of the residuals of the $\log R'_{HK}$ index
(see Fig.~\ref{fig:II}).

The magnetic cycle component in Eq.~(\ref{eq:model}) is thus given
in this simple model by
\begin{equation}
  \label{eq:magcycle}
  \mathrm{Mag.\ cycle}(t_k) = A \times \mathrm{smoothed\ activity}_k,
\end{equation}
where $A$ is the proportionality coefficient that is to be adjusted.
In order to have velocity units for $A$
and to minimize correlations between $A$ and $\gamma_i$,
we rescaled and centered the smoothed activity indicator
to have a zero mean and a semi-amplitude of one.

The magnetic cycle component of Eq.~(\ref{eq:magcycle}) might introduce an
offset in our model before and after the fiber change as a result
of the possible offset in the $\log R'_{HK}$ index.
This offset adds up to the offset in RV that is already considered in our model
(parameters $\gamma_i$ in Eq.~(\ref{eq:model})).
In the following results, the fitted offset ($\gamma_2-\gamma_1$)
is therefore always a combination of the RV offset and the $\log R'_{HK}$ offset.

Figure~\ref{fig:III} shows a periodogram (see Sect.~\ref{sec:perio})
of the RV data
with and without the magnetic cycle in our model.
When the magnetic cycle is ignored, the periodogram
is dominated by a peak around 3,000~d (see Fig.~\ref{fig:III}).
The period of this signal is compatible with the period
found in the $\log R'_{HK}$ index (see Fig.~\ref{fig:II}),
but the phase is opposite.
When the (negative) proportionality coefficient between the RVs
and the smoothed activity index is adjusted, the low-frequency part of
the periodogram ($P \gtrsim 100$ d) almost disappears (see Fig.~\ref{fig:III}).

\subsection{Short-term activity and correlated noise}
\label{sec:noise}

We modeled the short-term part of the activity as well as additional instrumental noise
(not included in the measurement uncertainties provided by the pipeline,
which only considers the photon noise contribution)
using a Gaussian process
\citep[e.g.,][]{baluev_orbital_2011,haywood_planets_2014}.
We assumed the noise ($\epsilon_k$ in Eq.~(\ref{eq:model}))
to follow a multivariate Gaussian random variable
with covariance matrix $C$ given by
\begin{equation}
  C_{i,j} = \delta_{i,j} (\sigma_i^2 + \sigma_J^2) + g(|t_j-t_i|),
\end{equation}
where $\sigma_i$ is the estimated error bar of the $i$-th measurement,
$\sigma_J$ is the stellar jitter,
$\delta_{i,j}$ is the Kronecker symbol, and
$g$ is the covariance function which is to be determined.
Several shapes for the covariance function have been proposed in the literature:
a squared exponential decay \citep[e.g.,][]{baluev_orbital_2011}
of amplitude $\sigma_G$ and correlation timescale ($\tau_G$)
\begin{equation}
  \label{eq:gfunc}
  g(\Delta t) = \sigma_G^2 \exp\left(-\frac{\Delta t^2}{2\tau_G^2}\right),
\end{equation}
or a more complex function
\citep[e.g.,][]{haywood_planets_2014}
that includes the rotation period of the star
($P_\mathrm{rot}$) and a smoothing parameter ($\eta$)
\begin{equation}
  g(\Delta t) = \sigma_G^2 \exp\left(
  -\frac{\Delta t^2}{2\tau_G^2}
  - \frac{2}{\eta^2} \sin^2\left(\frac{\pi\Delta t}{P_\mathrm{rot}}\right)
  \right).
\end{equation}
In the case of \object{HD~215152}, we find that these models give similar results.
In particular, in contrast to the case of \object{Corot-7} \citep{haywood_planets_2014},
the activity is weak and does not show significant correlation at the rotation
period of the star (i.e., around 42~d).
We thus use the simpler squared exponential decay function
(Eq.~(\ref{eq:gfunc})).
This function equals $\sigma_G^2$ on the diagonal ($\Delta t = 0$)
and rapidly decreases for $\Delta t \gg \tau_G$ (far from the diagonal).
In order to improve the computation efficiency, we neglected all the sub- and super-diagonals
for which all terms satisfy $g(\Delta t) < 10^{-3} \sigma_G^2$.
We thus obtained a band matrix (whose bandwidth depends on $\tau_G$)
that allowed us to use faster dedicated algorithms for the matrix inversion
and for computing the determinant.

For a given noise covariance matrix $C$ and given model parameters,
the likelihood function reads \citep[e.g.,][]{baluev_orbital_2011}
\begin{equation}
  \label{eq:likelihood}
  \log\mathcal{L} = -\frac{n}{2} \log(2\pi) - \frac{1}{2}\det C
  -\frac{1}{2} r^T C^{-1} r,
\end{equation}
where $r$ is the vector of the residuals.

\subsection{Periodograms}
\label{sec:perio}

\begin{figure}
  \centering
  \includegraphics[width=\linewidth]{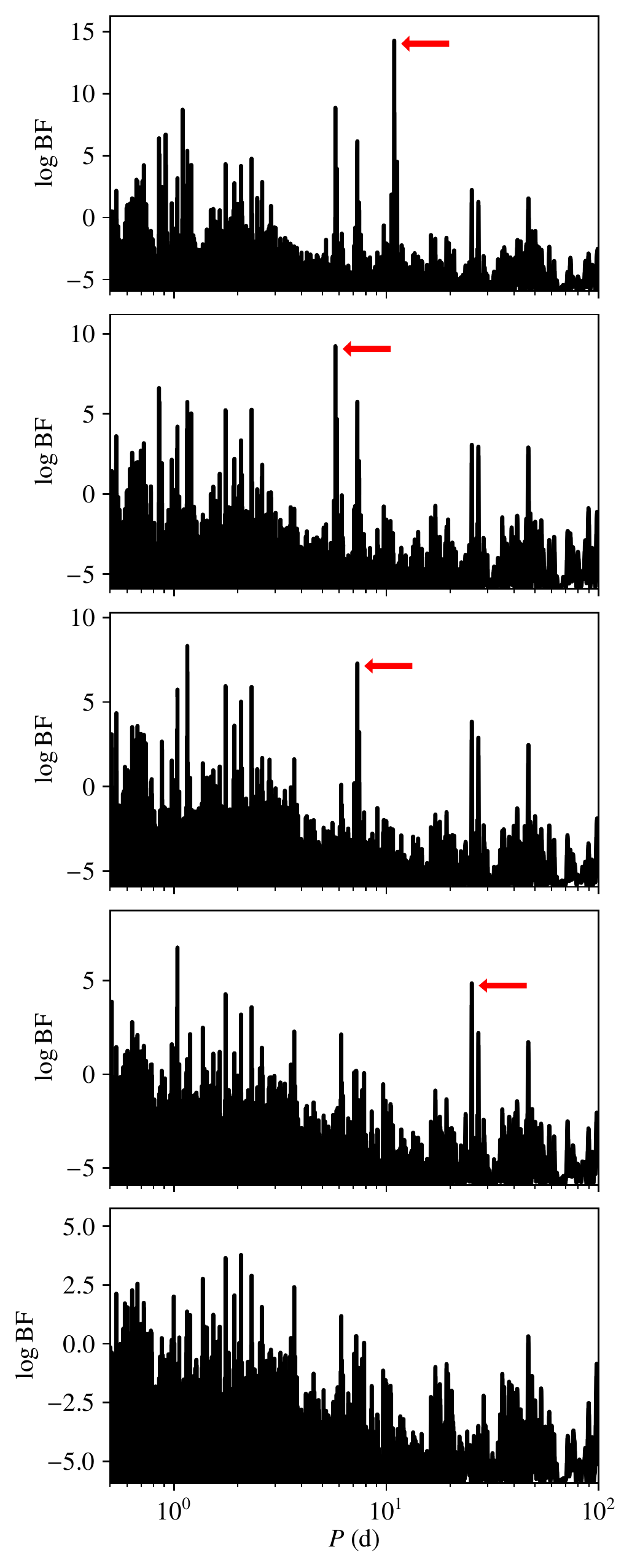}
  \caption{Periodogram of the RVs at each step of
    constructing our Keplerian model (see Sect.~\ref{sec:perio}).
    We highlight (\textit{red} arrows) the selected period (highest peak)
    for each planet: 10.87, 5.76, 7.28, and 25.2~d.
    We also show (\textit{bottom}) the periodogram of the residuals.
    Intermediate values of the planetary and noise parameters are
    given in Table~\ref{tab:II}.
  }
  \label{fig:IV}
\end{figure}

\begin{table*}
  \begin{center}
    \caption{Evolution of the noise and planetary parameters
      at each step of the construction of our Keplerian model (see Sect.~\ref{sec:perio} and Fig.~\ref{fig:IV}).}
    \begin{tabular}{c|cccll}
      \hline
      \hline
      number of planets & $\sigma_J$\ (m/s) & $\sigma_G$\ (m/s) & $\tau_G$\ (d) & $P$\ (d) & $K$\ (m/s)\\
      \hline
      0 & 0.000 & 1.799 & 1.035 & & \\
      1 & 0.000 & 1.645 & 0.966 & 10.866 & 1.069\\
      2 & 0.585 & 1.443 & 1.568 & 10.866, 5.760 & 1.019, 0.798\\
      3 & 0.653 & 1.327 & 1.939 & 10.865, 5.760, 7.282 & 0.970, 0.765, 0.666\\
      4 & 0.651 & 1.228 & 1.881 & 10.865, 5.760, 7.282, 25.197 & 0.963, 0.773, 0.675, 0.747\\
      \hline
    \end{tabular}
    \label{tab:II}
  \end{center}
\end{table*}

In order to determine the Keplerian component of our model (Eq.~\ref{eq:model}),
we started by inspecting the data using periodograms.
We sequentially added planets in the model.
At each step, we first subtracted the Keplerian part
of the last model (previously found planets) from the RVs.
For each period $P$ to be tested,
we then compared the model with the new planet (sinusoidal signal of period $P$)
with the model without it (reference).
For each period $P$ and for the reference,
we readjusted all other model parameters ($A$, $\gamma_i$, $\sigma_G$, and $\tau_G$),
except for the previously found planet parameters,
by maximizing the likelihood.
We then used the Bayesian information criterion
\citep[BIC, see][]{schwarz_estimating_1978}
\begin{equation}
  \mathrm{BIC} = 2 \log \mathcal{L}_\mathrm{max} - p \log n
\end{equation}
(where $p$ is the number of parameters of a given model)
to estimate the Bayes factor between the models
\begin{equation}
  \log \mathrm{BF}(P) = \frac{\mathrm{BIC}(P) - \mathrm{BIC_{ref}}}{2}.
\end{equation}
Following \citet{feng_agatha_2017}, we used these $\log \mathrm{BF}$ values
to draw the periodogram, and we used a threshold at $\log \mathrm{BF} = 5$ to
add a planet candidate in our model.
The main difference in our approach is that we modeled the correlated noise using
a Gaussian process (see Sect.~\ref{sec:noise}),
in contrast to \citet{feng_agatha_2017}, who used a moving-average model.
The complete model (including all previous planets)
was then readjusted by maximizing the likelihood
using a downhill simplex algorithm
\citep{nelder_simplex_1965},
and we iterated the process.
At this stage, we did not allow eccentric orbits.
The successive periodograms thus built are shown in Fig.~\ref{fig:IV}.
The evolution of the noise and planetary parameters at each step of the process
are detailed in Table~\ref{tab:II}.
We find four planet candidates at periods
10.87, 5.76, 7.28, and 25.2~d.

The aliases at one sidereal day (1.16 and 1.04~d)
of the two last candidates (7.28 and 25.2~d, respectively)
show slightly higher $\log \mathrm{BF}$ values than the periods we selected.
Given that we find a correlation timescale ($\tau_G$)
of about 1-2 days (see Table~\ref{tab:II}),
we suspect that periods of about one day are significantly
affected by the correlated noise,
and we considered the candidates at 7.28 and 25.2~d to be more likely
than their aliases at one day.
However, we cannot rule out any of the possibilities.
In order to break the alias at one sidereal day
and to better characterize the correlated noise,
we would need a dedicated observing strategy
with several points per night during several consecutive nights.
The signal-to-noise ratio (S/N) and the 43 nights with two points
are not sufficient in this case
to apply a dealiasing technique such as is described
in \citet{dawson_radial_2010}.
As in the case of the $\log R'_{HK}$ index, there does not seem to be
a significant signal at the rotation period of the star (about 42~d)
in the periodogram of the RV residuals
(see Fig.~\ref{fig:IV}).

We note that the \object{HD~215152} system was listed in the
statistical study of \citet{mayor_harps_2011},
with a detailed analysis yet to be published.
In this previous analysis, only two planets were detected
at 7.28 and 10.87~d.
We still find these two planets,
but we find two other candidates as well (at 5.76 and 25.2~d).
However,
the number of data points increased greatly
since 2011, from 171 to 373 points,
the time span increased from 8 to 13 years,
and the sampling is better (more nights with two points per night).
Since all the candidates have small amplitudes,
it is not surprising that some of them were missed in \citet{mayor_harps_2011}.
As a comparison, we reran our analysis, but using only the first 171 data points.
This analysis differs from the analysis used by \citet{mayor_harps_2011}
since we here included the stitching correction,
magnetic cycle modeling,
and a more complex noise model (Gaussian process).
As in \citet{mayor_harps_2011}, we find the candidates at
10.87 and 7.28~d (with a strong alias at 1.16~d).
The 5.76~d signal is just below the detection threshold ($\log \mathrm{BF} < 5$),
and we do not detect the 25.2~d signal.
This confirms that the huge observational effort employed since 2011
was essential to better characterize this system.

\subsection{Full model sampling}
\label{sec:sampling}

\begin{table}
  \begin{center}
    \caption{Priors used for the full model sampling.}
    \begin{tabular}{cc|c}
      \hline
      \hline
      Parameter & (units) &\\
      \hline
      $\sigma_J$ & $(\mathrm{m/s})$ & $|\mathcal{N}|(\mu=0,\ \sigma=5)$\\
      $\sigma_G$ & $(\mathrm{m/s})$ & $|\mathcal{N}|(\mu=0,\ \sigma=5)$\\
      $\tau_G$ & $(\mathrm{d})$ & $\log\mathcal{N}(\mu=0,\ \sigma=3)$\\
      $\gamma_i$ & $(\mathrm{m/s})$ & $\mathcal{U}(\min=-10^6,\ \max=10^6)$\\
      $A$ & $(\mathrm{m/s})$ & $\mathcal{U}(\min=-10^3,\ \max=10^3)$\\
      $P^{-1}$ & $(1/\mathrm{d})$ & $\mathcal{U}(\min=0,\ \max=10)$\\
      $K$ & $(\mathrm{m/s})$ & $\mathcal{U}(\min=0,\ \max=100)$\\
      $\lambda(0)$ & $(\mathrm{deg})$ & $\mathcal{U}(\min=0,\ \max=360)$\\
      $e$ & & $\mathcal{B}(\alpha=0.867,\ \beta=3.03)$\tablefootmark{1}\\
      $\omega$ & $(\mathrm{deg})$ & $\mathcal{U}(\min=0,\ \max=360)$\\
      \hline
    \end{tabular}
    \tablefoot{
      \tablefoottext{1}{\citet{kipping_parametrizing_2013}}.
    }
    \label{tab:III}
  \end{center}
\end{table}

\begin{table*}
  \begin{center}
    \caption{Results of the elliptical fit (see Sect.~\ref{sec:sampling}).
      Values correspond to the ML, and errors give the 2$\sigma$ interval.}
    \begin{tabular}{cc|cccc}
      \hline
      \hline
      Parameter & (units) & & & &\\
      \hline
      $\sigma_J$ & $(\mathrm{m/s})$ & $0.632_{-0.595}^{+0.189}$ & & & \\
      $\sigma_G$ & $(\mathrm{m/s})$ & $1.262_{-0.145}^{+0.303}$ & & & \\
      $\tau_G$ & $(\mathrm{d})$ & $1.975_{-1.067}^{+0.551}$ & & & \\
      $\gamma_1$ & $(\mathrm{m/s})$ & $-13687.434_{-0.372}^{+0.175}$ & & & \\
      $\gamma_2$ & $(\mathrm{m/s})$ & $-13673.120_{-0.518}^{+0.901}$ & & & \\
      $A$ & $(\mathrm{m/s})$ & $-1.504_{-0.508}^{+0.291}$ & & & \\
      \hline
      planet & & b & c & d & e\\
      $P$ & $(\mathrm{d})$ & $5.75942_{-0.00095}^{+0.00190}$ & $7.28345_{-0.00869}^{+0.00343}$ & $10.86420_{-0.00500}^{+0.00617}$ & $25.1968_{-0.0526}^{+0.0478}$\\
      $K$ & $(\mathrm{m/s})$ & $0.907_{-0.364}^{+0.151}$ & $0.608_{-0.221}^{+0.315}$ & $0.986_{-0.317}^{+0.300}$ & $0.914_{-0.559}^{+0.128}$\\
      $\lambda(0)$ & $(\mathrm{deg})$ & $24.61_{-18.46}^{+28.11}$ & $-158.04_{-34.51}^{+30.20}$ & $98.24_{-20.07}^{+28.83}$ & $-37.80_{-33.37}^{+42.04}$\\
      $e\cos\omega$ & & $0.354_{-0.417}^{+0.094}$ & $-0.098_{-0.137}^{+0.402}$ & $-0.332_{-0.153}^{+0.440}$ & $-0.167_{-0.200}^{+0.370}$\\
      $e\sin\omega$ & & $-0.045_{-0.135}^{+0.236}$ & $-0.131_{-0.151}^{+0.276}$ & $-0.024_{-0.203}^{+0.214}$ & $0.048_{-0.264}^{+0.286}$\\
      \hline
      $e$ & & $0.357_{-0.351}^{+0.108}$ & $0.163_{-0.160}^{+0.223}$ & $0.333_{-0.326}^{+0.182}$ & $0.173_{-0.169}^{+0.287}$\\
      $\omega$ & $(\mathrm{deg})$ & $-7.3_{-143.4}^{+143.2}$ & $-126.9_{-164.7}^{+167.6}$ & $-175.8_{-131.6}^{+151.5}$ & $163.9_{-167.1}^{+167.3}$\\
      $m\sin i$ & $(m_\oplus)$ & $1.996_{-0.746}^{+0.414}$ & $1.526_{-0.561}^{+0.792}$ & $2.705_{-0.822}^{+0.911}$ & $3.467_{-2.133}^{+0.462}$\\
      $a$ & $(\mathrm{AU})$ & $0.057635_{-0.000755}^{+0.000742}$ & $0.067399_{-0.000898}^{+0.000854}$ & $0.08799_{-0.00115}^{+0.00113}$ & $0.15417_{-0.00204}^{+0.00199}$\\
      \hline
    \end{tabular}
    \label{tab:IV}
  \end{center}
\end{table*}

\begin{table*}
  \begin{center}
    \caption{Same as Table~\ref{tab:IV}, but for the circular fit.}
    \begin{tabular}{cc|cccc}
      \hline
      \hline
      Parameter & (units) & & & &\\
      \hline
      $\sigma_J$ & $(\mathrm{m/s})$ & $0.651_{-0.616}^{+0.176}$ & & & \\
      $\sigma_G$ & $(\mathrm{m/s})$ & $1.228_{-0.103}^{+0.338}$ & & & \\
      $\tau_G$ & $(\mathrm{d})$ & $1.881_{-0.988}^{+0.604}$ & & & \\
      $\gamma_1$ & $(\mathrm{m/s})$ & $-13687.517_{-0.274}^{+0.269}$ & & & \\
      $\gamma_2$ & $(\mathrm{m/s})$ & $-13672.936_{-0.730}^{+0.675}$ & & & \\
      $A$ & $(\mathrm{m/s})$ & $-1.612_{-0.399}^{+0.380}$ & & & \\
      \hline
      planet & & b & c & d & e\\
      $P$ & $(\mathrm{d})$ & $5.75999_{-0.00175}^{+0.00157}$ & $7.28243_{-0.00827}^{+0.00451}$ & $10.86499_{-0.00613}^{+0.00564}$ & $25.1967_{-0.0505}^{+0.0476}$\\
      $K$ & $(\mathrm{m/s})$ & $0.772_{-0.266}^{+0.211}$ & $0.675_{-0.284}^{+0.242}$ & $0.963_{-0.317}^{+0.277}$ & $0.747_{-0.384}^{+0.275}$\\
      $\lambda(0)$ & $(\mathrm{deg})$ & $30.43_{-25.49}^{+25.25}$ & $-156.03_{-37.66}^{+28.51}$ & $104.66_{-26.37}^{+23.52}$ & $-31.73_{-37.96}^{+35.00}$\\
      \hline
      $m\sin i$ & $(m_\oplus)$ & $1.819_{-0.629}^{+0.501}$ & $1.720_{-0.725}^{+0.618}$ & $2.801_{-0.923}^{+0.809}$ & $2.877_{-1.481}^{+1.063}$\\
      $a$ & $(\mathrm{AU})$ & $0.057638_{-0.000759}^{+0.000739}$ & $0.067393_{-0.000893}^{+0.000860}$ & $0.08799_{-0.00116}^{+0.00113}$ & $0.15417_{-0.00204}^{+0.00199}$\\
      \hline
    \end{tabular}
    \label{tab:V}
  \end{center}
\end{table*}

\begin{figure}
  \centering
  \includegraphics[width=\linewidth]{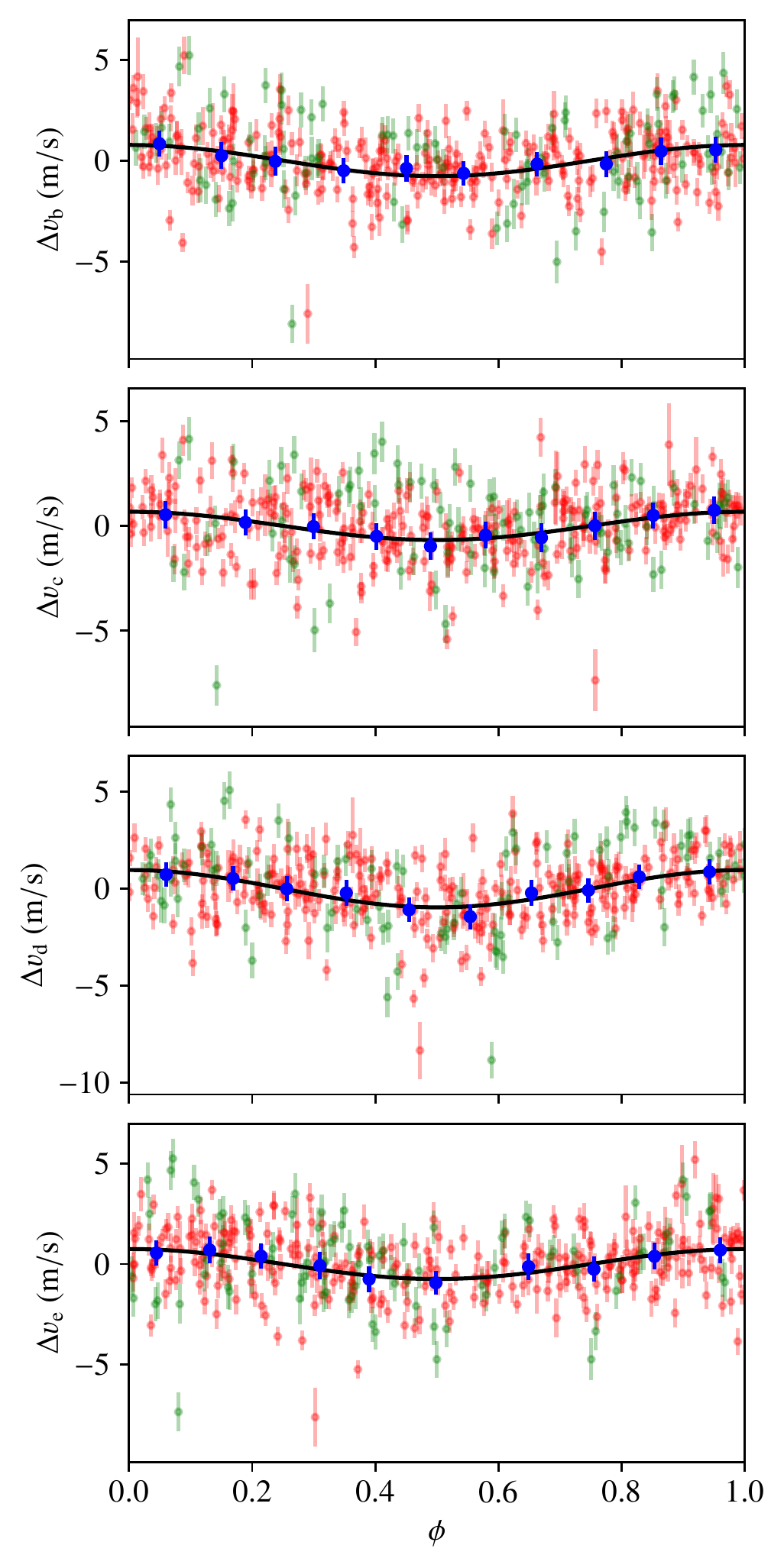}
  \caption{Phase-folded curve for each of the four planets obtained
    using the parameters of the circular ML.
    The \textit{black} line shows the model.
    \textit{Red} (\textit{green}) dots correspond to HARPS data
    taken before (after) the fiber upgrade.
    \textit{Blue} dots show the data binned in phase.
    For each panel, we subtracted the Keplerian signal
    associated with all other planets from the data,
    as well as the magnetic cycle component and instrumental offsets.}
  \label{fig:V}
\end{figure}

In order to better determine the orbital architecture of the system,
we explored the parameter space of our full model
(four planets, offsets, magnetic cycle, and Gaussian process)
using an adaptive Metropolis approach.
The sampling algorithm is described in Appendix~\ref{sec:amalgo},
and the priors are detailed in Table~\ref{tab:III}.
We ran two experiments, one with elliptical orbits and one with circular orbits.
In both cases, we ran the algorithm
for 2,000,000 steps, but only keep the last 1,500,000
steps to compute statistics.
Table~\ref{tab:IV} shows the results of the elliptical fit,
while the results of the circular fit are summarized in Table~\ref{tab:V}.
In both tables,
the uncertainty on the mass of the star (see Table~\ref{tab:I})
is taken into account to derive
the minimum masses ($m\sin i$)
and semi-major axes ($a$) of the planets.
Table~\ref{tab:IV} clearly shows that the eccentricities cannot be constrained with the current data.
This is expected for such low-amplitude signals
($K\lesssim 1$~m/s for all four planets).
However, all the other parameters remain compatible between the circular and the elliptic fit.
Moreover, since the whole system is very compact,
we can expect tidal dissipation to have severely damped any initial eccentricity.
The circular fit (Table~\ref{tab:V}) therefore probably represents the system better than the elliptical fit.
In the following, we use the set of parameters
corresponding to the best circular fit (i.e., maximum likelihood, ML).
We show the phase-folded RV curves for each
planet in Fig.~\ref{fig:V} as obtained with these parameters.

\section{System dynamics and stability}
\label{sec:dynamics}

\begin{figure}
  \centering
  \includegraphics[width=\linewidth]{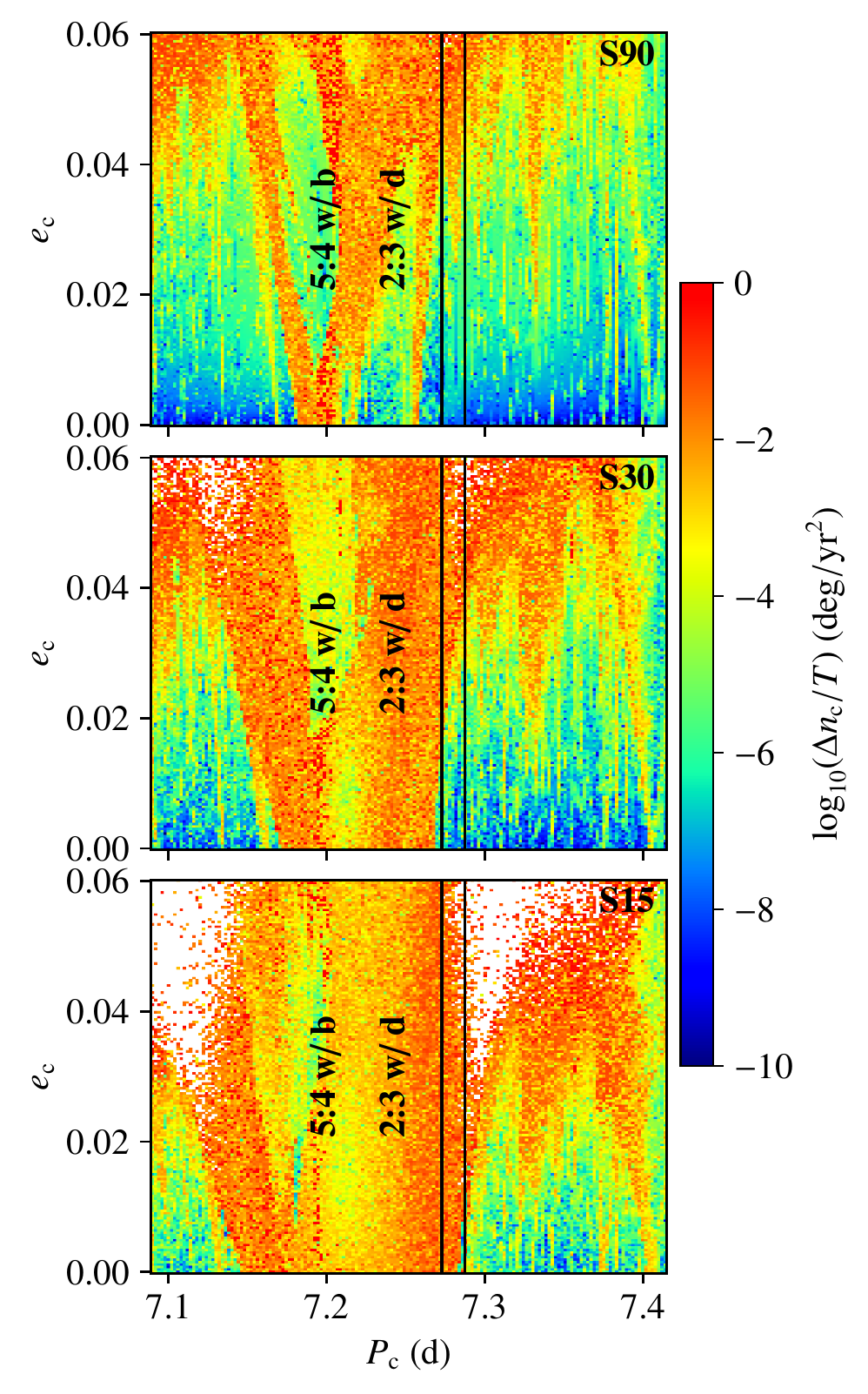}
  \caption{Stability analysis of \object{HD~215152}~c around the circular ML solution (see Table~\ref{tab:V})
    and assuming planetary inclinations of
    $90^\circ$ (\textit{top}), $30^\circ$ (\textit{center}),
    and $15^\circ$ (\textit{bottom}).
    Blue points correspond to stable orbits,
    red points to unstable orbits,
    and white points correspond to highly unstable solutions
    (collision or ejection of a planet during the 1~kyr of the simulation).
    The two vertical black lines highlight the 2$\sigma$ interval for
    the period of Planet~c.
    We also highlight the location of the 5:4 resonance with Planet~b and the 3:2 resonance with Planet~d.}
  \label{fig:VI}
\end{figure}

The planetary system around \object{HD~215152} is very compact, especially the three inner planets, which exhibit period ratios of $P_c/P_b = 1.26$ (close to 5/4) and $P_d/P_c = 1.49$ (close to 3/2).
This compact architecture raises several interesting questions regarding the dynamics of the system,
such as whether the planets are locked in mean-motion resonances,
if the observed system is stable,
and how the stability of the system depends on the eccentricities and the inclinations of the planets.
In order to partially answer these questions, we ran three sets of N-body integrations of the system.
We used the ML solution of the circular model (see Table~\ref{tab:V})
as a reference and varied the period and eccentricity of Planet~c
to generate three grids of 161x161 (25921) initial conditions.
In the first set of simulations (referred to as S90),
we assumed the four planets to be coplanar,
with inclinations of $90^\circ$ ($m = m \sin i$).
In the second set of simulations (S30),
we assumed the four planets to be coplanar,
with inclinations of $30^\circ$ ($m = 2 m \sin i$).
Finally, the third set (S15) corresponds to inclinations of $15^\circ$
($m \approx 3.86 m\sin i$).

For each initial condition, we integrated the system for 1~kyr using GENGA \citep{grimm_genga_2014},
and we assessed the stability of the system using the NAFF stability indicator \citep{laskar_chaotic_1990,laskar_frequency_1993,correia_coralie_2005}.
We show the results of these integrations in Fig.~\ref{fig:VI}.
In the S90 simulations,
where the masses are equal to minimum masses,
the system is stable at low eccentricities
($e_c \lesssim 0.03$, see Fig.~\ref{fig:VI} \textit{top}).
If the masses are twice the minimum masses
(S30, Fig.~\ref{fig:VI} \textit{center}),
the system can still be stable, but only at lower eccentricities ($e_c \lesssim 0.02$).
Finally, for the S15 simulations
(Fig.~\ref{fig:VI} \textit{bottom}),
the system is unstable even at zero eccentricities.
We also note in Fig.~\ref{fig:VI}
that stable orbits that are compatible with the data
correspond to a system that clearly lies outside both
the 5:4 resonance between Planets~b and~c
and the 3:2 resonance between Planets~c and~d.

These simulations illustrate that constraints on the eccentricities and
the inclinations of the planets could be derived from studying the
system stability.
However, we note that the bounds we obtain on the eccentricity of Planet~c
should be taken with caution since we only varied a few orbital parameters
here (the period and eccentricity of Planet~c,
and the overall system inclination).
In order to obtain more precise constraints, all the orbital parameters of all the planets need to be let free to vary.

\section{Discussion}
\label{sec:discussion}

We reported the discovery of a compact system of four super-Earth planets around
\object{HD~215152}.
The planets have periods of 5.76, 7.28, 10.86, and 25.2~d,
and minimum masses of 1.8, 1.7, 2.8, and 2.9~$M_\oplus$ (respectively).
Planets~b and~c are close to but outside of a 5:4 mean motion resonance
(period ratio of 1.26),
and Planets~c and d are close to but outside of a 3:2 mean motion resonance (period ratio of 1.49).
Owing to the low masses of the planets (low S/N), it is very difficult
to constrain the planets eccentricities.
However, using N-body simulations, we showed that interesting constraints on the eccentricities and on the inclinations of the planets
could be derived from studying the system stability.
Assuming inclinations of $90^\circ$ for all the planets, we find that the
system remains stable for $e_c \lesssim 0.03$.
If the planets remain coplanar but have inclinations of $30^\circ$
(masses of twice the minimum masses), the maximum eccentricity is about 0.02.
Moreover, for a coplanar system with an inclination of $15^\circ$,
the system is unstable even at zero eccentricities.
Interestingly, these values ($e_c\lesssim 0.03$)
are on the order of magnitude of eccentricities found
in \textit{Kepler} multi-transiting systems that are close to resonances
\citep[eccentricities inferred using transit timing variations, see][]{wu_density_2013}.
This is not surprising as the \object{HD~215152} planetary system
shares many properties with \textit{Kepler} compact multi-planetary systems.
In addition, low eccentricities are expected in such a compact system
because of the tides exerted by the star on the planets.
We note that the bounds we obtain on the eccentricity of Planet~c should only
be taken as very crude estimates because we did not explore the
entire parameter space and only varied a few orbital parameters.
A more complete dynamical analysis is thus necessary to obtain precise constraints
on the planet eccentricities and inclinations,
but this is beyond the scope of this paper.

\begin{acknowledgements}
  We thank the anonymous referee for his/her useful comments.
  We acknowledge financial support from the Swiss National Science Foundation (SNSF).
  This work has in part been carried out within the framework of
  the National Centre for Competence in Research PlanetS
  supported by SNSF.
  XD is grateful to The Branco Weiss Fellowship--Society in Science
  for its financial support.
  PF and NCS acknowledge support by Funda\c{c}\~ao para a Ci\^encia e a Tecnologia (FCT) through Investigador FCT contracts of reference IF/01037/2013/CP1191/CT0001 and IF/00169/2012/CP0150/CT0002, respectively, and POPH/FSE (EC) by FEDER funding through the program ``Programa Operacional de Factores de Competitividade - COMPETE''. PF further acknowledges support from Funda\c{c}\~ao para a Ci\^encia e a Tecnologia (FCT) in the form of an exploratory project of reference IF/01037/2013/CP1191/CT0001.
  RA acknowledges the Spanish Ministry of Economy and Competitiveness (MINECO) for the financial support under the Ramón y Cajal program RYC-2010-06519, and the program RETOS ESP2014-57495-C2-1-R and ESP2016-80435-C2-2-R.
\end{acknowledgements}

\bibliographystyle{aa}
\bibliography{hd215152.bib}

\appendix

\section{Adaptive Metropolis algorithm}
\label{sec:amalgo}

In order to explore the parameter space of our model,
we used an adaptive Metropolis algorithm inspired by
\citet{haario_adaptive_2001}.
This algorithm is similar to the classical
Metropolis-Hastings algorithm
\citep{metropolis_equation_1953,hastings_generalization_1970},
except that at each step the proposal is randomly generated
around the previous point, using the covariance matrix
of all preceding points \citep[see][]{haario_adaptive_2001}.

We additionally included in the algorithm a possibility to go back
around an anterior point (not the previous one).
This idea is inspired by the tuned jump proposal described in
\citet{farr_more_2014},
and it allowed us to improve the mixing of the chain.
We implemented it as follows:
\begin{itemize}
  \item At each step, there is a given probability
  ($p_\mathrm{go-back} = 0.01$ in our case)
  to make a go-back proposal.
  Otherwise, a classical proposal around the previous point is made.
  \item In case of a go-back proposal, an anterior point is randomly selected,
  and a proposal is made around this point
  (still using the covariance matrix of all previous steps).
  \item Since the proposal is not symmetrical, and in order to maintain detailed balance, we computed the forward and backward jump probabilities \citep[e.g.,][]{farr_more_2014} and took them into account in computing the acceptance probability.
  \item The computation of forward and backward jump probabilities becomes very expensive in computer time when the number of possible anterior points increases. We thus only considered one point over hundred as a possible go-back point (around which the proposal can be made).
\end{itemize}

\begin{figure}
  \centering
  \includegraphics[width=\linewidth]{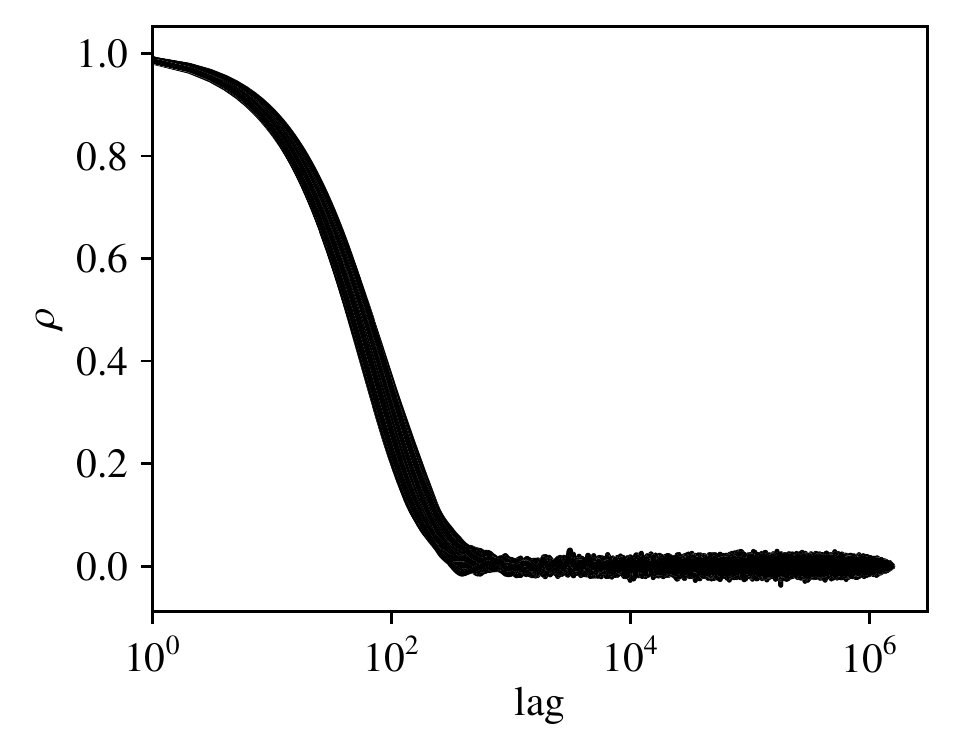}
  \caption{Autocorrelation functions for the 26 parameters
    of the elliptical model, the log likelihood, and log posterior,
    computed on the last 1,500,000 steps of the Metropolis algorithm.
    The maximum integrated autocorrelation time is 223,
    which corresponds to an effective sample size of 6,721.}
  \label{fig:AI}
\end{figure}

\begin{figure}
  \centering
  \includegraphics[width=\linewidth]{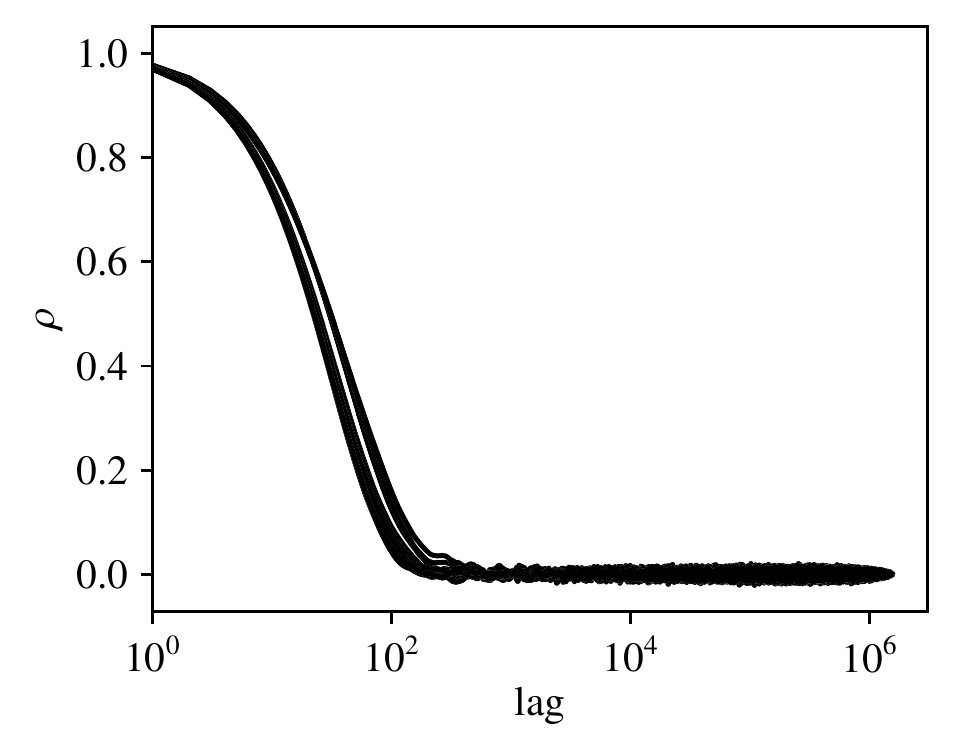}
  \caption{Autocorrelation functions for the 18 parameters
    of the circular model, the log likelihood, and log posterior,
    computed on the last 1,500,000 steps of the Metropolis algorithm.
    The maximum integrated autocorrelation time is 112,
    which corresponds to an effective sample size of 13,417.}
  \label{fig:AII}
\end{figure}

We used this algorithm to explore the parameter space
for both the elliptical case and the circular case.
The results are detailed in Sect.~\ref{sec:sampling}
(Tables~\ref{tab:IV} and \ref{tab:V}).
In order to illustrate the mixing properties of our chains,
we show in Figs.~\ref{fig:AI} and~\ref{fig:AII} the autocorrelation functions
of all parameters in both cases.
In the elliptical case, the maximum integrated autocorrelation time is 223
(effective sample size of 6,721).
This is sufficient to estimate $2\sigma$ error bars.
In the circular case, the mixing is even better (integrated autocorrelation time of 112, and effective sample size of 13,417).
This is not surprising since the problem has fewer dimensions and is much better constrained.

\end{document}